\documentclass[fleqn,twoside]{article}
\usepackage{espcrc2}

\usepackage{graphicx,epsfig}
\usepackage[figuresright]{rotating}

\newcommand{\AmS}{{\protect\the\textfont2
  A\kern-.1667em\lower.5ex\hbox{M}\kern-.125emS}}

\newcommand{\MeV}{\ensuremath{\mathrm{MeV}}}
\newcommand{\mm}{\ensuremath{\mathrm{mm}}}
\newcommand{\m}{\ensuremath{\mathrm{m}}}
\newcommand{\cm}{\ensuremath{\mathrm{cm}}}
\newcommand{\uus}{\ensuremath{\mu{\mathrm{s}}}}
\newcommand{\micron}{\ensuremath{\mathrm{\mu m}}}
\newcommand{\ponn}{\ensuremath{\mathrm{{\it p^{+}}-on-{\it n}}}}
\newcommand{\nonn}{\ensuremath{\mathrm{{\it n^{+}}-on-{\it n}}}}
\newcommand{\Neq}{\ensuremath{\mathrm{n_{eq}}}}
\newcommand{\dgr}{\ensuremath{^\circ}}
\title{Silicon Detectors at the LHC}

\author{C. Parkes\address[UG]{Department of Physics and Astronomy, 
        Kelvin Building, \\ 
        University of Glasgow, Glasgow, G12 8QQ, Scotland, U.K.}%
        \thanks{email: Chris.Parkes@cern.ch}
       }
       
\begin{document}

\begin{abstract}
The silicon systems of the Large Hadron Collider (LHC) detectors are briefly described. The complexity and diversity of the projects are illustrated by highlighting for discussion different components of the silicon systems in each experiment.

\vspace{1pc}
\end{abstract}

\maketitle

\section{Introduction}
The LHC detectors will have to function in a high radiation environment, fluences of up to $10^{15}$~1~\MeV\ neutron equivalents (\Neq)$/\cm^2$ are expected over a 10 year period. They will have to cope with a 40 MHz collision rate and high particle multiplicities. In this environment, the detectors must provide efficient tracking and b-tagging capabilities.

These requirements are met by the silicon systems that are being constructed  for the LHC detectors on a scale previously unprecedented in particle physics.

\section{ALICE}

ALICE, the heavy ion collider experiment of the LHC, has the highest expected particle multiplicities at the LHC: up to 8000 charged particles per unit of rapidity. In order to achieve a good two-track separation three dimensional hit information is required wherever possible. 

The primary functions of the ALICE silicon system~\cite{bib:ALICE} are the determination of primary and secondary vertices necessary for the reconstruction of charm and hyperon decays, to provide identification (using dE/dx) for low momentum particles and to assist the ALICE time projection chamber in particle tracking.

The silicon system consists of six cylindrical layers of silicon detectors. The two inner layers, at radii of $4-7~\cm$, and coping with multiplicities of up to 90~particles/\cm$^2$ are comprised of 240~modules containing 9.8~million $50~\times~425~\micron$ \ponn\ pixels .

The two central layers consist of silicon drift detectors, ALICE is the only experiment at the LHC to exploit this technology. The need for three dimensional information combined with the relatively modest central event rate ($\sim$40~Hz trigger) and radiation hardness requirements (12.7~krad in 10 years for the inner layer) at ALICE make this a suitable technology choice. A series of ${ \it p^+}$ field strips are implanted on both sides of an {\it n}-type silicon wafer and connected by built-in voltage dividers. Hence, under an applied bias voltage, an electric drift field parallel to the wafer surface is produced. Electrons liberated by charged particles are drifted up to 35 \mm\ in 4.5 \uus\ and collected on the $256~{\it n^+}$ anodes. The doping inhomogeneities in the neutron transmutation doped material cause the electrons to deviate significantly from their ideal trajectory and this must be compensated for. The deep sub-micron CMOS readout electronics samples the collected charge at 40 MHz with a 10 bit ADC. After corrections are applied a spatial precision of 25 \micron\ in the drift co-ordinate and 33 \micron\ in the perpendicular co-ordinate has been achieved over the full drift distance of 35 \mm. 

The outermost two layers of the silicon system, at radii of 39 and 43 \cm, are comprised of 1770 modules containing 2.7~million 95~\micron\ pitch double sided silicon strips. Analogue read-out, implemented in the 0.25~\micron\ CMOS process,  is used in order that the dE/dx information is preserved. 

\section{LHCb}

LHCb, the dedicated b physics experiment at the LHC, has a forward geometry covering the range 15 to 300 mrad. The tracking system comprises a set of circular silicon sensors known as the vertex locator~\cite{bib:VELO}, a silicon strip station and three hybrid tracking stations. The silicon sensors will all be read out with radiation hard analogue electronics.
 
The hybrid tracking stations employ two different technologies: they have silicon sensors in the higher occupancy region close to the beam and straw drift tubes surrounding these. The silicon strip station and the hybrid layers will use the same sensor design and a similar layout. Each station comprises four layersof single sided 200~\micron\ \ponn\ pitch sensors. Two of these layers are rotated by a small stereo angle. Typically, two 11cm long sensors are connected together to form a ladder.   

\begin{figure}[htb]
\begin{center}
\epsfig{file=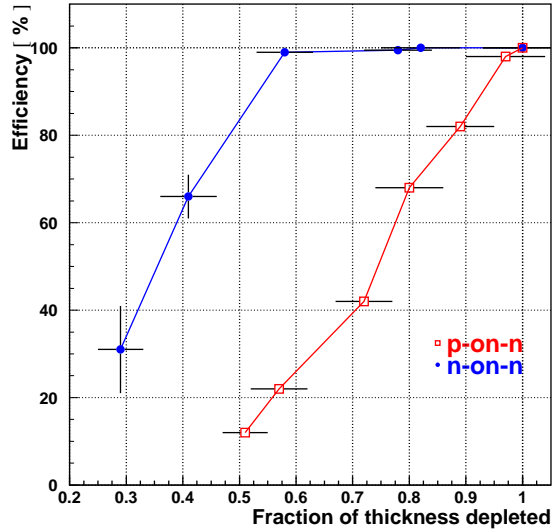, width=0.5\textwidth}
\caption{A comparison of the performance of irradiated LHCb VELO \ponn\ and \nonn\ prototype sensors. \label{fig:velo}}
\end{center}
\end{figure}

The vertex locator (VELO) is required to perform primary and b decay vertex reconstruction, with the displaced vertex capabilities being used in the second level trigger. Hence, good vertex resolution and fast tracking capabilities were strong considerations for the design. The VELO uses two sets of 42 4.2~\cm\ outer radius semi-circular sensors, arranged over approximately one metre with the beam direction as their axis, covering the interaction point and the region downstream. The sensors are of two types, having strips that run either radially or circumferentially with pitches varying between 40 and 100 \micron. The two sets of half disks are separated by 6~\cm\ during injection of the LHC beam, but the active strips at the inner radius of the disks are brought to only 8~\mm\ from the LHC beam when stable conditions are provided. Naturally this leads to an extreme radiation environment, a fluence of $10^{14}$~\Neq/$\cm^2$ per year is anticipated.  The performance of irradiated \ponn\ and \nonn\ prototype sensors, evaluated in a test beam, are compared in Figure \ref{fig:velo}. \nonn\ sensors have been chosen as they remain fully efficient even when only 60~\% of the detector thickness is depleted. A best resolution of 4~\micron\ has been achieved with these sensors.

\section{ATLAS}

The inner detector tracking system of ATLAS~\cite{bib:ATLAS} consists of pixels, silicon strips and a straw tracker.

The pixel system has three forward disks per side and three barrel layers at radii of 5 to 12~\cm. The detector is composed of 1500 modules and has 100 million $50 \times 400~\micron^2$ \nonn\ pixels. The modules are mounted on carbon frames, where their positions must be monitored and maintained at a 10~\micron\ accuracy. The 15~kW of power dissipated in the $1.5~\m^2$ pixel system is removed by a coolant in the integrated cooling ducts. A new 7~\m\ long insertable tubular layout has been designed which allows independent installation and removal of the pixels from the silicon strip layers. In addition to decoupling the construction schedule for the pixels and strips this will also facilitate upgrades, for example a change in the inner pixel layer which (due to radiation damage) is only anticipated to be operational for the first three to five years of the LHC. The pixels are readout with a In/SnPb bump-bonded front-end chip implemented in the radiation hard 0.25~\micron\ CMOS process. A recent production run has been highly successful with 85~\% of the chips passing the initial tests, and irradiated bump-bonded assemblies retaining high efficiency (98\%), low noise (350~$e^-$) and the expected resolution of 13~\micron. 

Four barrel layers at radii of 27 to 52~\cm\ and nine end cap disks per side of silicon strips surround the pixel detector. Single sided \ponn\ strips are used throughout with a square sensor containing 80~\micron\ pitch strips used in the barrel and wedge geometries used in the forward region.  The sensors are readout with binary DMILL BiCMOS front-end chips. A module consists of two sets, mounted back to back, of two daisy-chained single sided sensors with a 40~mrad stereo angle between the planes. The effective strip length is around 12~\cm.

The designs of all the LHC detectors are now well advanced and indeed production of the larger systems of ATLAS and CMS has already commenced. Indeed, in the case of ATLAS, 85~\% of the silicon strip sensors have already been delivered, with less than 1~\% of the manufacturer's supplied detectors failing the ATLAS tests. Module production is also now underway.

In order to successfully assemble such large systems a well structured testing and quality assurance plan is necessary. Extensive tests are performed on the components such as the sensors, hybrids and front-end chips before assembly, during assembly and on the completed modules. The scale of these tests is illustrated by listing the tests that will be performed on the final assembled ATLAS silicon module: 100 points are measured on each detector; Leakage current measurements up to 500V are taken and the stability checked over a 24 hour time period; a range of electrical tests are performed including the requirement that 99~\% of the strips are efficient; the modules are required to withstand ten thermal cycles between -25\dgr C and +40\dgr C; finally the modules are operated, clocked and triggered,  at the standard operating environment and temperature of -7\dgr C for 24 hours.

\section{CMS}

\begin{figure*}[htb]
\begin{center}
\mbox{\epsfig{file=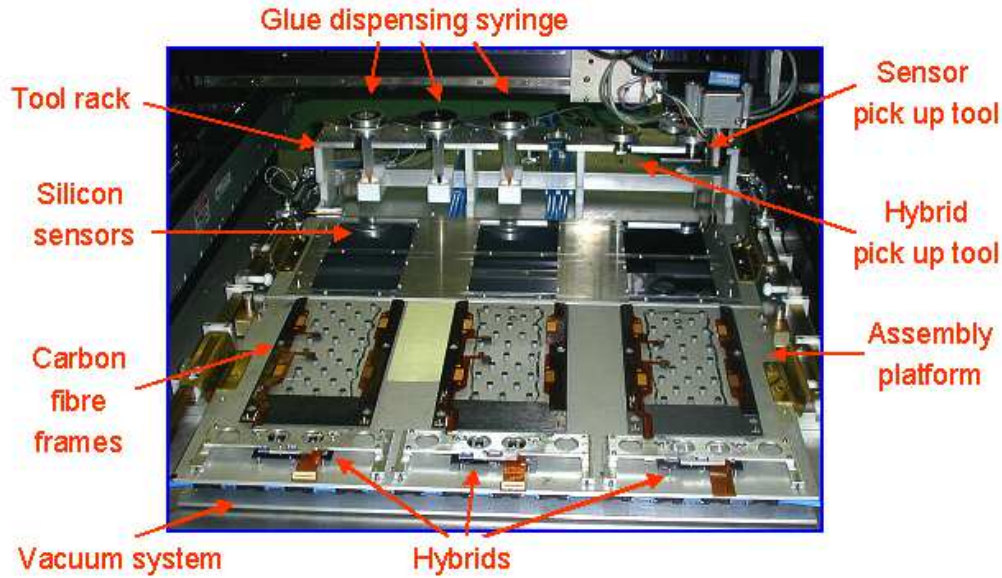, width=0.85\textwidth,}}

\caption{The features of the CMS automatic module assembly system are indicated.\label{fig:CMSassembly}}
\end{center}
\end{figure*}

CMS has by far the largest of the silicon systems, as they have adopted an entirely silicon solution to their tracking \cite{bib:CMS}. 

The initial barrel pixel system consists of two layers close to the interaction point at radii of 4.4 and 7.2~\cm\ and two forward disks, with a $100 \times 150 \micron^2$ pixel size. The inner layer is expected to receive a fluence of about $3 \times 10^{14}\Neq/\cm^2$ per year at full machine luminosity, with the readout chip being the most vulnerable part of the system and expected to withstand approximately two years of running. At a fluence of  $6 \times 10^{14}~\Neq/\cm^2$ the sensors are expected to be operated at less than half the full depletion voltage, hence the choice of \nonn\ pixels.

One of the key differences of the CMS and ATLAS silicon strip detectors has already been mentioned: while the ATLAS system has 4,000 modules the CMS system has a staggering 15,000 modules to assemble. There are 10 detector layers in the barrel region and three inner disks and nine outer disks per side of the detector:  a total of 206~$m^2$ of silicon. The CMS module design uses either two or four \ponn\ sensors in a module for single sided or back-to-back layers. The strip pitches vary between 61 \micron\ in the inner barrel layer at 21~\cm\ and  140 \micron\ in the outer layer at 115~\cm.

Another significant difference is the choice of readout mode for the system. While ATLAS has selected the cheaper option of binary readout, the additional control and resolution of analogue readout has been chosen by CMS and the other two experiments. The larger data volume of the CMS analogue system is then readout of the 128 channel front-end chips and passed to a laser driver which transmits the signals in fibre optic cables over a 100\m\ distance to the counting room. 

An entirely automated system has been developed both for the module assembly and for the testing. The setup that is used at each of the seven laboratories assembling modules is shown in  figure \ref{fig:CMSassembly}. The system uses video from a CCD video camera to automatically locate fiducial marks on each of the module components. Glue is dispensed onto the carbon fibre frame and then the silicon sensors and hybrid are positioned by a robotic arm. Results show that two sensors in a module have an rms translational displacement of only 3~\micron\ and a rotational accuracy of 1.5~mrad has been achieved. Three modules can be assembled simultaneously in around 25 minutes.

\section{Closing Remarks}
With the development and construction of the LHC a paradigm shift in silicon detector history is occurring. There are nine distinct subsystems, over $280~m^2$ of silicon and 22,000 modules to assemble. The silicon systems of the LHC do not simply provide superior vertexing capabilities or provide the first space points for tracking, like their LEP forerunners, but rather their combined attributes of spatial precision, speed and radiation tolerance have gained these technologies their roles as primary components of the tracking system for the LHC experiments.

\section{Acknowledgements}
I would like to thank the many members of the silicon groups of the four LHC experiments who assisted in the preparation of my presentation and of this paper, particularly Ariella Cattai, Leonardo Rossi and Flavio Tosello.

\end{document}